\documentclass[journal,twoside,web]{ieeecolor}
\usepackage{generic}
\usepackage{cite}
\usepackage{amsmath,amssymb,amsfonts}
\usepackage{algorithmic}
\usepackage{graphicx}
\usepackage{algorithm,algorithmic}
\usepackage{array}
\usepackage{graphicx}
\usepackage{stfloats}
\usepackage{textcomp}
\usepackage{soul}

\def\BibTeX{{\rm B\kern-.05em{\sc i\kern-.025em b}\kern-.08em
    T\kern-.1667em\lower.7ex\hbox{E}\kern-.125emX}}
\begin{document}
\onecolumn
\begin{center}
{\huge  \bf Deep Representation Learning-Based Dynamic Trajectory Phenotyping for Acute Respiratory Failure in Medical Intensive Care Units \\ \vspace{2cm}

{\LARGE \it Graphical Abstract}}
\end{center}

\begin{figure*}[h]
    \vspace{1cm}
    \includegraphics[width=0.98\linewidth]{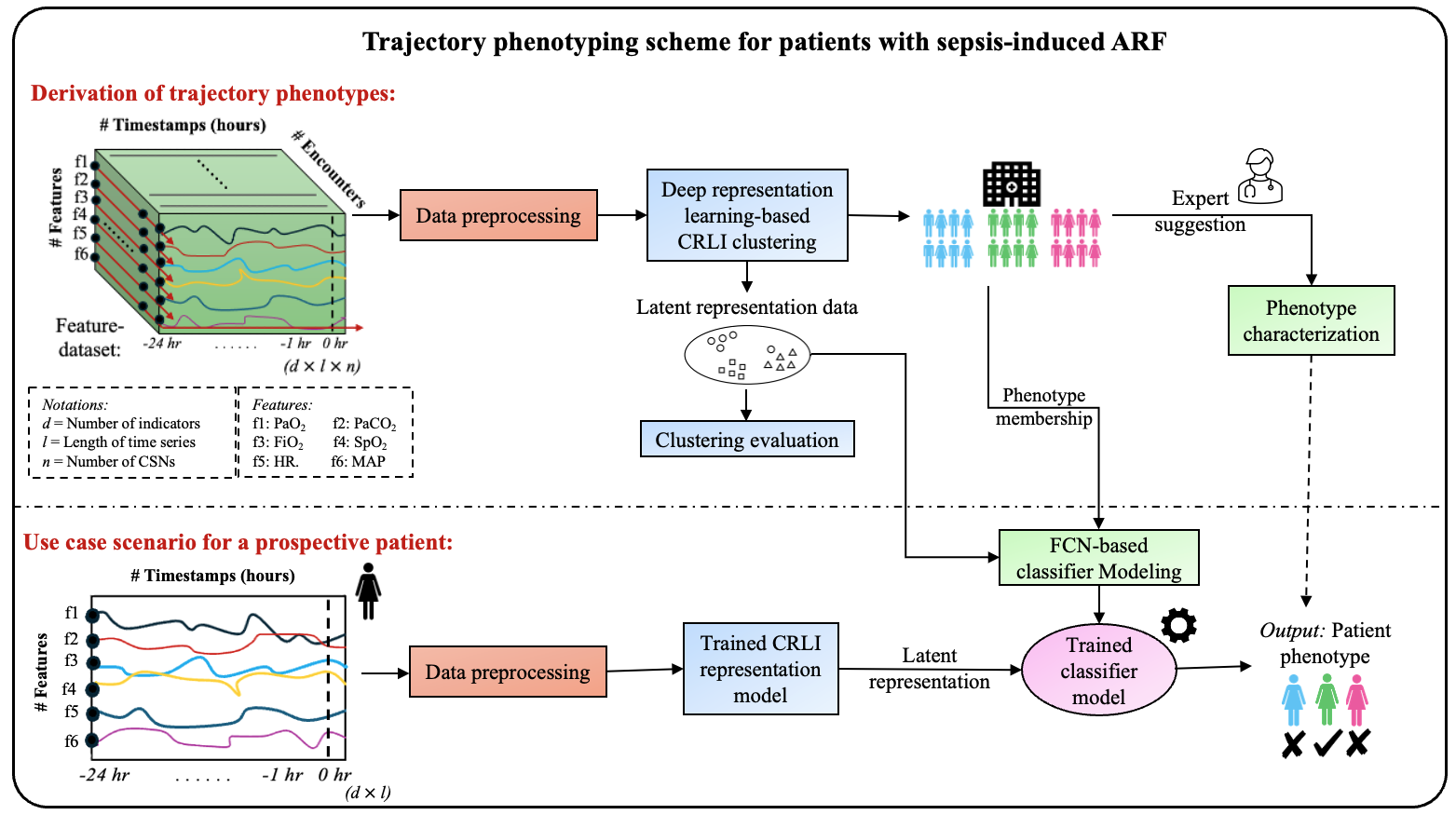}\par
    \vspace{0.5cm}
    {{\bf GRAPHICAL ABSTRACT:} Schematic showing our proposed method for the derivation of novel trajectory phenotypes of patients in medical ICU with sepsis-induced acute respiratory failure (ARF) using a deep representation learning framework on multivariate time series data along with a prospective use-case scenario in clinical settings. The illustration covers: (A) Process for deriving trajectory phenotypes. The model was paired with a parsimonious set of routinely collected cardio-respiratory variables. After data preprocessing and representation learning, our method yielded four data-driven distinct phenotypes that were further characterized by a critical care expert. (B) A prospective use-case scenario that covers supervised algorithm development and application on prospective and/or hidden datasets to determine generalizable phenotype membership. This may help in effective prognosis, tailored treatment regimens, and overall clinical management.     
}
    \label{fig:graphical_abstract}
\end{figure*}
\twocolumn
%%%%%%%%%%%%%%%%%%%%%%%%%%%%%%%%%%%%%%%%%%%%%%%%%%%%%

%\title{Dynamic Trajectory Phenotyping for Acute Respiratory Failure}
\title{Deep Representation Learning-Based Dynamic Trajectory Phenotyping for Acute Respiratory Failure in Medical Intensive Care Units}
\author{Alan Wu, Tilendra Choudhary, Pulakesh Upadhyaya, 
Ayman Ali, Philip Yang, and Rishikesan Kamaleswaran 
\thanks{The authors were supported by the National Institutes of Health under Award Numbers R01GM139967 and R21GM151703. (Corresponding author: Tilendra Choudhary, e-mail: tilendra.choudhary@duke.edu)} %(Corresponding author: Alan Wu, e-mail: alan.wu@emory.edu)}
\thanks{Alan Wu is with the Department of Quantitative Theory and Methods, Emory University, Atlanta, GA, USA. Tilendra Choudhary, Pulakesh Upadhyaya, Ayman Ali, and Rishikesan Kamaleswaran are with the Duke University School of Medicine, Durham, NC, USA. Philip Yang is with the Division of Pulmonary, Allergy, Critical Care, and Sleep Medicine, Emory University, Atlanta, GA, USA.}}
% \thanks{Pulakesh Upadhyaya, Duke University School of Medicine, Durham, NC, USA.}
% \thanks{Ayman Ali, Duke University School of Medicine, Durham, NC, USA.}
% \thanks{Philip Yang, Division of Pulmonary, Allergy, Critical Care, and Sleep Medicine, Emory University, Atlanta, GA, USA.}
% \thanks{Rishikesan Kamaleswaran, Duke University School of Medicine, Durham, NC, USA.}}
% (e-mail: tilendra.choudhary@duke.edu)
% (e-mail: pulakesh.upadhyaya@duke.edu)
% (email: philip.yang@emory.edu)
% (e-mail: r.kamaleswaran@duke.edu)

\maketitle

\begin{abstract}
    Sepsis-induced acute respiratory failure (ARF) is a serious complication with a poor prognosis. This paper presents a deep representation learning-based phenotyping method to identify distinct groups of clinical trajectories of septic patients with ARF. For this retrospective study, we created a dataset from electronic medical records (EMR) consisting of data from sepsis patients admitted to medical intensive care units who required at least 24 hours of invasive mechanical ventilation at a quarternary care academic hospital in southeast USA for the years 2016--2021. A total of N=3349 patient encounters were included in this study. Clustering Representation Learning on Incomplete Time Series Data (CRLI) algorithm was applied to a parsimonious set of EMR variables in this data set. To validate the optimal number of clusters, the K-means algorithm was used in conjunction with dynamic time warping. Our model yielded four distinct patient phenotypes that were characterized as liver dysfunction/heterogeneous, hypercapnia, hypoxemia, and multiple organ dysfunction syndrome by a critical care expert. A Kaplan-Meier analysis to compare the 28-day mortality trends exhibited significant differences (p$<$ 0.005) between the four phenotypes.  The study demonstrates the utility of our deep representation learning-based approach in unraveling phenotypes that reflect the heterogeneity in sepsis-induced ARF in terms of different mortality outcomes and severity. These phenotypes might reveal important clinical insights into an effective prognosis and tailored treatment strategies.
\end{abstract}


\section{Introduction}
\IEEEPARstart{S}{epsis} is a serious syndrome defined as life-threatening organ dysfunction caused by dysregulated patient response to infection\cite{singer2016third}. This syndrome results in more than 350,000 deaths in America and around 11 million worldwide each year\cite{cdcSepsisBodys,whoSepsis}. Patients with sepsis in intensive care units (ICU) often develop acute respiratory failure (ARF), in which the respiratory system fails to adequately oxygenate and/or remove carbon dioxide from the blood, which, in severe cases, requires invasive mechanical ventilation to maintain end-organ perfusion\cite{GURKA2008773, mirabile2023respiratory}. Previous studies of patients with sepsis and ARF requiring invasive mechanical ventilation (IMV) have demonstrated a poor prognosis with mortality rates of approximately 43\%, which is a source of significant concern for critical care experts \cite{lewandowski1995incidence}. Respiratory failure is a widely present complication in patients with sepsis, with various determinants, physiological processes, and immunological responses, resulting in extreme heterogeneity in clinical trajectories involving multiorgan dysfunctions and other comorbidities that are difficult to interpret and characterize in clinical settings. Therefore, in order to study the clinical course of ARF, it becomes imperative to differentiate between various phenotypes of patients, each of which may follow similar clinical trajectories.

Trajectory clustering is a domain where similar trajectories in time or space are grouped into distinct clusters. Since clinical trajectories used in this study are multivariate time series data, ``time series" and ``trajectory" are used interchangeably in this paper. In the existing literature, many methods have been reported to cluster time series data. One group of methods utilizes distance or similarity metrics such as Euclidean distance or dynamic time warping (DTW) distance to calculate the distance between time series \cite{muller2007dynamic}. A traditional clustering technique, e.g., K-means, DBSCAN, hierarchical clustering, is then used to cluster the distance matrix. Another group of methods utilizes the recent path-breaking advances in the field of representation learning to design deep neural networks to cluster time series. They do not require that a distance metric be applied to the data before fitting the model. Examples include Clustering Representation Learning on Incomplete Time Series Data (CRLI)\cite{Ma_Chen_Li_Cottrell_2021}, Variational Deep Embedding with Recurrence (VaDER)\cite{de2019deep}, and Deep Temporal Contrastive Clustering (DTCC)\cite{zhong2023deep}. These methods utilize various neural network frameworks, such as convolutional neural networks, autoencoders, and recurrent neural networks. They also may employ a clustering objective such as K-means to encourage the development of clusters~\cite{alqahtani2021deep}. 

The variability of clinical signs and symptoms leading to sepsis-induced ARF can make the task of patient management particularly challenging. Although clinicians have access to a large variety of lab tests and vital signs, the breadth of pathophysiological etiologies that can lead to sepsis-induced ARF (such as localized direct lung injury from pneumonia, inflammatory lung injury from systemic inflammation) could result in difficulty in predicting the clinical trajectory of an individual patient after developing ARF. Trajectory clustering may help simplify the task of prediction by grouping patients into subgroups that follow a similar path prior to ventilation. Trajectory-based phenotyping methods can also help clinicians create more targeted and patient-specific treatment regimens. These subgroups can reveal opportunities to identify patients early based on their initial disease progression and also allow clinicians to intervene prior to the need for invasive respiratory support.

\begin{figure*}[t]
\centering
    \includegraphics[width=0.92\linewidth]{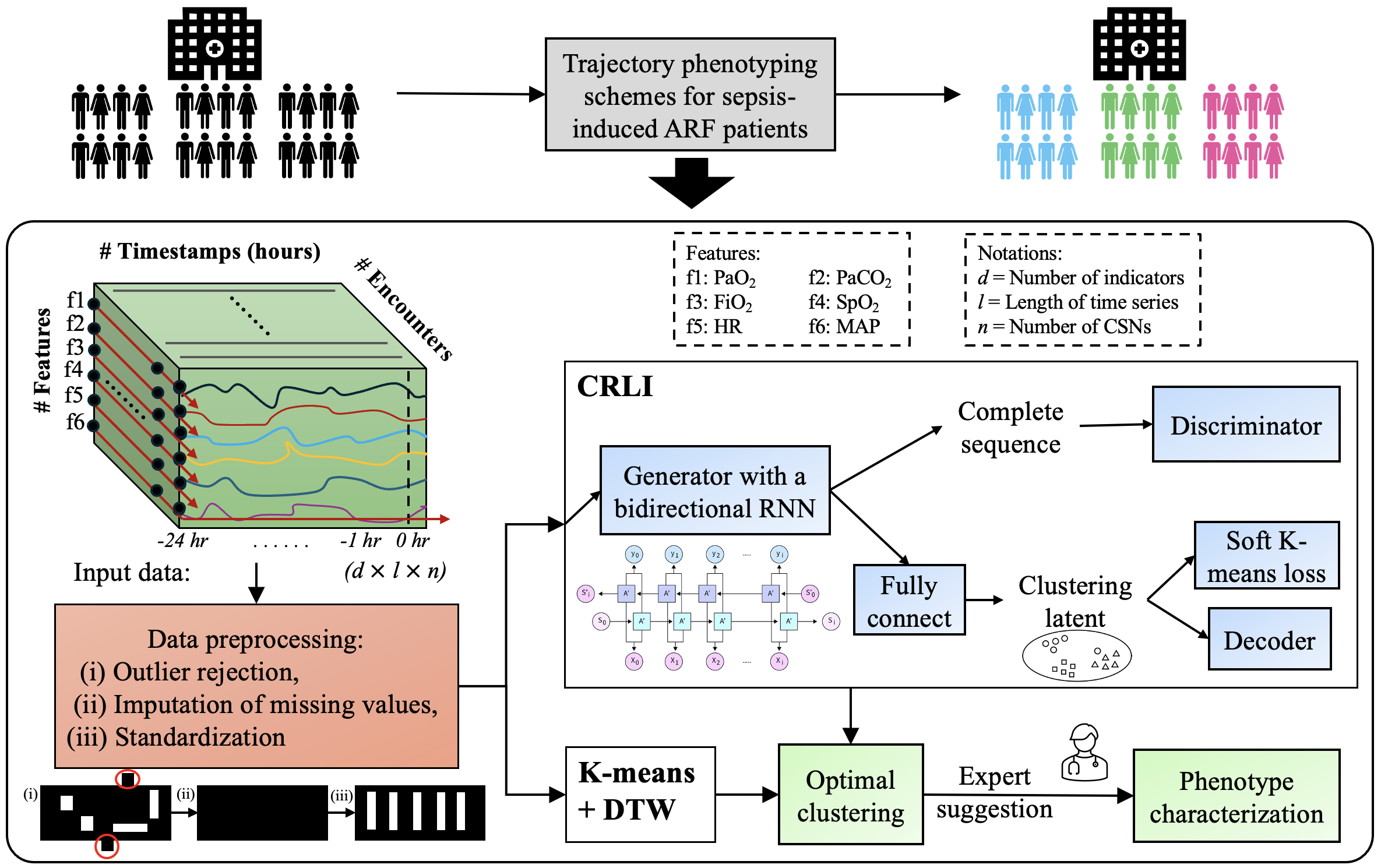}
    \caption{Block diagram depicting an overview of our study design for sepsis-induced ARF trajectory phenotyping.}
    \label{fig:overview}
\end{figure*}

Trajectory clustering has been used in many areas of medicine\cite{schulam2015clustering,liu2015temporal}. For example, a previous work has used longitudinal data to find respiratory subphenotypes for COVID-19-related acute respiratory distress syndrome (ARDS)\cite{bos2021longitudinal}. However, the time series data of a more generic cohort of sepsis-induced ARF patients prior to ventilation (as opposed to a specific complication like ARDS) has not been analyzed yet using such methodologies. It is commonly observed that once patients with sepsis are mechanically ventilated, there is a high associated mortality and multi-organ failures, and grouping patients with similar clinical trajectories  may help optimizing therapies and improving such poor outcomes \cite{komorowski2022sepsis}. 

Furthermore, the large-scale archival and storage of time-dependent patient information in electronic medical records (EMR) has facilitated the use of novel deep learning techniques in various clinical tasks \cite{8086133}. Thus, in this paper, we proposed an ARF trajectory phenotyping method using a deep representation learning-based approach on multivariate time series EMR data to elucidate novel phenotypes of sepsis-induced ARF in ICU and characterize them with the help of a clinician expert. The rest of the paper is organized as follows. In Section II, our study design and method are presented. Section III presents phenotyping results and analysis, followed by a discussion section in IV. Finally, conclusions are drawn in Section V.

\section{Methodology}

A block diagram, representing the study design of the proposed sepsis-induced ARF trajectory phenotyping method, is shown in Fig.~\ref{fig:overview}. It consists of four major blocks: data collection, data preprocessing, trajectory clustering, and phenotype characterization. Data collection involves patient selection and collection of EMR data up to 24 hours prior to ventilation. Data preprocessing introduces a multivariate feature-set from a parsimonious set of cardio-respiratory features, and data operations such as outlier rejection, missingness imputation and standardization. Subsequently, a deep representation learning-based trajectory clustering is used to derive optimal data-driven phenotypes that are characterized by following expert suggestions. 
The various components of the study are described in detail in subsequent subsections.

\subsection{Data Collection}
This study is a retrospective cohort study conducted at Emory University Hospital network, located in Georgia, United States. This study was reviewed and approved by the Emory University IRB (\#STUDY00000302) and performed by following the ethical standards of the Emory University IRB and the Helsinki Declaration of 1975. 
Adult patients ($\geq$18 years) admitted to medical intensive care units (MICU) with sepsis (according to the sepsis-3 criteria \cite{singer2016third}) between 2016--2021 and developed ARF during their hospital stay in the ICU were included. ARF was defined as the requirement of $\geq$24 hours of invasive mechanical ventilation (IMV).

For elucidating the trajectory phenotypes early in the course of sepsis-induced ARF, we considered up to 24 hours of data prior to the time of IMV. Clinical data collected from the EMR, including laboratory values and vital signs, were used for phenotyping. A set of demographic factors (such as age, sex, race and ethnicity), mortality outcomes, and comorbidity information were also incorporated for further analysis of the derived phenotypes. We defined time of IMV as the moment at which the mechanical ventilation parameters (positive end-expiratory pressure (PEEP), tidal volume, and/or plateau pressure) were first ever recorded in the EMR in patients who followed the aforementioned inclusion criteria. We excluded patients who did not meet sepsis-3 criteria, patients admitted to neurological ICUs or surgical ICUs, or those whose EMR did not include any physiological data up to 24 hours preceding the IMV.

\subsection{Data Preprocessing}
We used a parsimonious set of six cardio-respiratory features, namely partial pressure of oxygen (PaO$_2$),  partial pressure of carbon dioxide (PaCO$_2$), fraction of inspired oxygen (FiO$_2$), oxygen saturation by pulse oximetry (SpO$_2$), heart rate (HR) and mean arterial pressure (MAP),  as inputs to the clustering algorithms. Other demographic and routinely collected clinical features, comorbidities at admission, as well as outcomes such as mortality were analyzed after clustering. Values of different variables were deemed to be spurious if they were above or below the accepted range. The spurious values were dropped and corrected. The following three imputation steps were performed subsequently for all the features: 
\begin{itemize}
    \item \textit{Forward filling:} The missing values were imputed with the last known value in the time series.
    \item \textit{Backward filling:} The missing values were imputed with the next known value in the time series.
    \item \textit{Median filling:} Completely missing time series of any specific feature of a patient was filled with the population median. However, for FiO$_2$, 0.21 was used to replace such missingness in time series. A value of 0.21 was used as replacement because the room air contains 21\% of oxygen. 
    %For all other indicators, the median value prior to imputation was used.
\end{itemize}
   
In order to make the data fixed length for our algorithms, the first value of the time series was added to the beginning to make the time series 24 data-points long. Each indicator was standardized using the formula: $z_i = (f_i - \mu) / \sigma$; where $f_i$ is the unstandardized feature, and $\mu$ and $\sigma$ denote mean and standard deviation of $f_i$, respectively. The order of patient encounters was then randomized.

%A new MAP indicator was created by imputing the arterial line MAP indicator with cuff-based MAP values, and then imputing the remaining missing values with the formula $ \frac{2}{3} \times $ cuff-based diastolic blood pressure + $\frac{1}{3} \times$ cuff-based systolic blood pressure. Cuff-based MAP and arterial line MAP were then removed from the dataset. A new S/F ratio indicator was calculated by dividing SpO$_2$ by FiO$_2$. A new P/F ratio indicator was calculated by dividing PaO$_2$ by FiO$_2$. A total of 38 indicators remained. 

\subsection{Clustering Algorithms}
\subsubsection{Clustering Representation Learning on Incomplete time-series data (CRLI)}
The state-of-the-art clustering algorithm, CRLI, developed in 2023, is a deep learning neural network used for clustering multivariate time series data~\cite{Ma_Chen_Li_Cottrell_2021}. 
The architecture is shown in Figure~\ref{fig:overview}. It consists of a generative adversarial network (GAN) framework with a bidirectional recurrent neural network (RNN) and an encoder-decoder network with a soft K-means objective. The clustering process relies on the deep temporal representations learned by the architecture.
The overall loss function of CRLI is:
$$L_{CRLI} = L_{pre} + L_{rec} + L_{adv} +\lambda \cdot L_{K-means}$$ 

\noindent
The loss $L_{pre}$ is derived from the bidirectional RNN of the generator and attempts to capture the temporal dynamics of the time series data. $L_{rec}$ is the reconstruction loss of the autoencoder network and attempts to identify informative features. $L_{adv}$ is the adversarial loss and attempts to capture the error propagation during clustering. $L_{K-means}$ is the soft K-means loss and encourages the network to form clusters. $\lambda$ is the weight of the K-means loss\cite{Ma_Chen_Li_Cottrell_2021}.
%2 generator layers, a recurrent neural network (RNN) size of 256, gated recurrent unit (GRU) as the RNN cell type, an output dimension of 256 and 128 for the layers in the fully-connected network of the decoder 
 CRLI consists of a bidirectional multi-layer RNN as generator (encoder), a single-layer RNN as decoder and five-layer RNN as discriminator. Each of these RNNs is a gated recurrent unit (GRU). We used a weight of $10^{-3}$ for the K-means loss, a batch size equal to the size of the entire dataset, 100,000 epochs, and a patience of 100 epochs for early stopping\cite{du2023pypots}. The default values of all other hyper-parameters were used. 
For the implementation of CRLI, the python package Pypots was used\cite{du2023pypots}. 

\subsubsection{K-means and Dynamic Time Warping (DTW)}
The second clustering algorithm utilizes dynamic time warping (DTW) and K-means. DTW is a technique to determine the similarity between two time series. DTW allows for the warping of the time series along the time dimension to better match two time series. This algorithm is useful because similar trajectories often vary along the time dimension. 
The DTW distance is defined as:
$DTW(X,Y) = \min\{C_p(X, Y) | \textrm{$p$ is a warping path}\}$
in which $X = (x_1, x_2,\cdots, x_N)$ and $Y = (y_1, y_2,\cdots, y_M)$ are time series, $C_p(X, Y)$ is the total cost function: the sum of absolute differences between $X$ and $Y$ using the index pairs given by $p$, and $p = (p_1, p_2,\cdots, p_L)$. $p$ is defined as a warping path when $p_l = (n_l, m_l) \in [1 : N] \times [1 : M]$ for $l \in [1 : L = \max(N,M)]$ such that boundary, montonicity, and step size conditions hold.
In summary, these conditions mean that the boundary paths are between the the boundary indexes, the paths between the indexes do not cross, \emph{i.e.}, if $a>b$ and $c>d$, then $p_i = (n_a, m_d)$ and $p_j = (n_b, m_c)$ do not coexist, and that all indexes are assigned to an opposing index\cite{muller2007dynamic}. 

K-means is a common clustering algorithm that creates $k$ clusters by minimizing the distance between the points and the cluster center. While it is computationally efficient and versatile in its ability to utilize different types of distance measures, it has a few limitations. It requires the number of clusters to be set prior to its execution, it may converge to local minima instead of the global minimum, and it assumes that the clusters are spherical.

K-means was run with 10 random initiations, 1000 max iterations, and the metric DTW for 3, 4, and 5 clusters.  The implementation from the python library Tslearn was used since it conveniently packages K-means and DTW together in the model TimeSeriesKMeans\cite{tavenard2020tslearn}.

\subsection{Optimal Clustering and Phenotype Characterization}
The performance of the models and the selection of the clusters was evaluated using silhouette scores. For CRLI, silhouette scores were calculated on an intermediate, latent representation called clustering latent. This is the 2D representation of the time series that the neural network generates. For K-means + DTW, silhouette scores will be calculated using DTW as a distance metric. The silhouette score will be calculated for each predefined number of clusters: 3, 4, and 5. A higher silhouette score suggests better clustering.
The silhouette score is defined as the average silhouette width for all points. The silhouette width is defined as:
$$s(x_i) = (b(x_i)-a(x_i))/\max\{b(x_i), a(x_i)\}$$

\noindent
where $x_i$ is an element in cluster $k$, $a(x_i)$ is the average distance of $x_i$ to all other elements in the cluster $k$, and $b(x_i)$ is the minimum average distance from $x_i$ to all points in a cluster other than $k$\cite{shutaywi2021silhouette}. The silhouette score was calculated using implementations from Tslearn and Pypots\cite{tavenard2020tslearn,du2023pypots}.
The clustered trajectories were then assessed by an expert critical care physician for characterizing and naming the phenotypes. 

\subsection{Statistical Analysis}
A Kaplan-Meier curve shows the survival probability overtime of a group of patients~\cite{goel2010understanding}. The statistical differences in mortality trends between clusters were evaluated using Kaplan-Meier curves and a multivariate log-rank test \cite{lifelinesLifelinesx2014}. One way analysis of variance tests (ANOVA) were also performed on comorbidity indexes created from comorbidity data.

\section{Results}
\subsection{Demographics}
The details of the demographics, including race, ethnicity, sex, and age of our patient cohort are listed in Table \ref{tab:demographics_pat_pop}. Our data consists of 3,225 patients with a total of 3,349 patient encounters or admissions. Out of the 3,225 patients, 1295 patients died with an in-hospital mortality rate of 40.2\%. 
\begin{table}[t]
\caption{Demographics of patient population.}
    \centering
    
    \begin{tabular}{|c|c|}
    \hline 
    \textbf{Demographic Feature} & \textbf{ Value} \\ 
    \hline 
         Age mean (standard deviation) &   62.3 (15.5)  \\ 
Males, count (\%)  & 1814 (54.2) \\
Race: African American or Black, count (\%) & 1624 (48.5) \\
Race: Caucasian or White, count (\%) & 1442 (43.1) \\
Ethnicity: Hispanic, count (\%)  & 129 (3.9) \\
Ethnicity: Non-Hispanic, count (\%)  & 2975 (88.8) \\
\hline 
    \end{tabular}
    
    \label{tab:demographics_pat_pop}
\end{table}
\subsection{Silhouette Score Analysis}

The silhouette score for CRLI was the highest for 4 clusters, denoting the optimal cluster selection. The choice of 4 clusters was also validated by the highest silhouette score from the K-means + DTW algorithm.  While the silhouette scores may not be comparable between clustering methods, according to Ma \textit{et al.}, CRLI performed better than K-means + DTW after analysis using more reliable ground truth based metrics \cite{Ma_Chen_Li_Cottrell_2021}.  The silhouette scores of the two methods are shown in Table \ref{tab:silhouette}. 

\begin{table}[t]
\caption {Silhouette scores calculated for CRLI using clustering latent and silhouette scores calculated for K-means + DTW using DTW.}
    \centering

    \begin{tabular}{|c|>{\centering\arraybackslash}p{0.25\linewidth}|>{\centering\arraybackslash}p{0.25\linewidth}|}
    \hline 
    \textbf{Clusters}& \textbf{CRLI} &\textbf{K-means + DTW}\\ 
\hline 
3 & 0.362  &0.177 \\ 
4 & 0.389  &0.193 \\ 
5 & 0.206  &0.153 \\ 
\hline 
    \end{tabular}
    
    \label{tab:silhouette}
\end{table}

\subsection{Phenotyping Results}
By utilizing the silhouette score criterion, the CRLI algorithm yielded four distinct data-driven clusters.  
Various demographics and mortality outcomes of each cluster are shown in Table~\ref{tab:demographics}. Both the number of unique patients and encounters were included because some patients had more than one encounter. Derived clusters were analyzed and characterized by an expert physician into phenotypes. Cluster 1 was characterized as a liver dysfunction/heterogeneous phenotype, cluster 2 as a hypercania phenotype, cluster 3 as a hypoxemia phenotype, and cluster 4 as a multiple organ dysfunction syndrome (MODS) phenotype.

\begin{figure*}[t]
    \centering
    \includegraphics[width=0.87\linewidth]{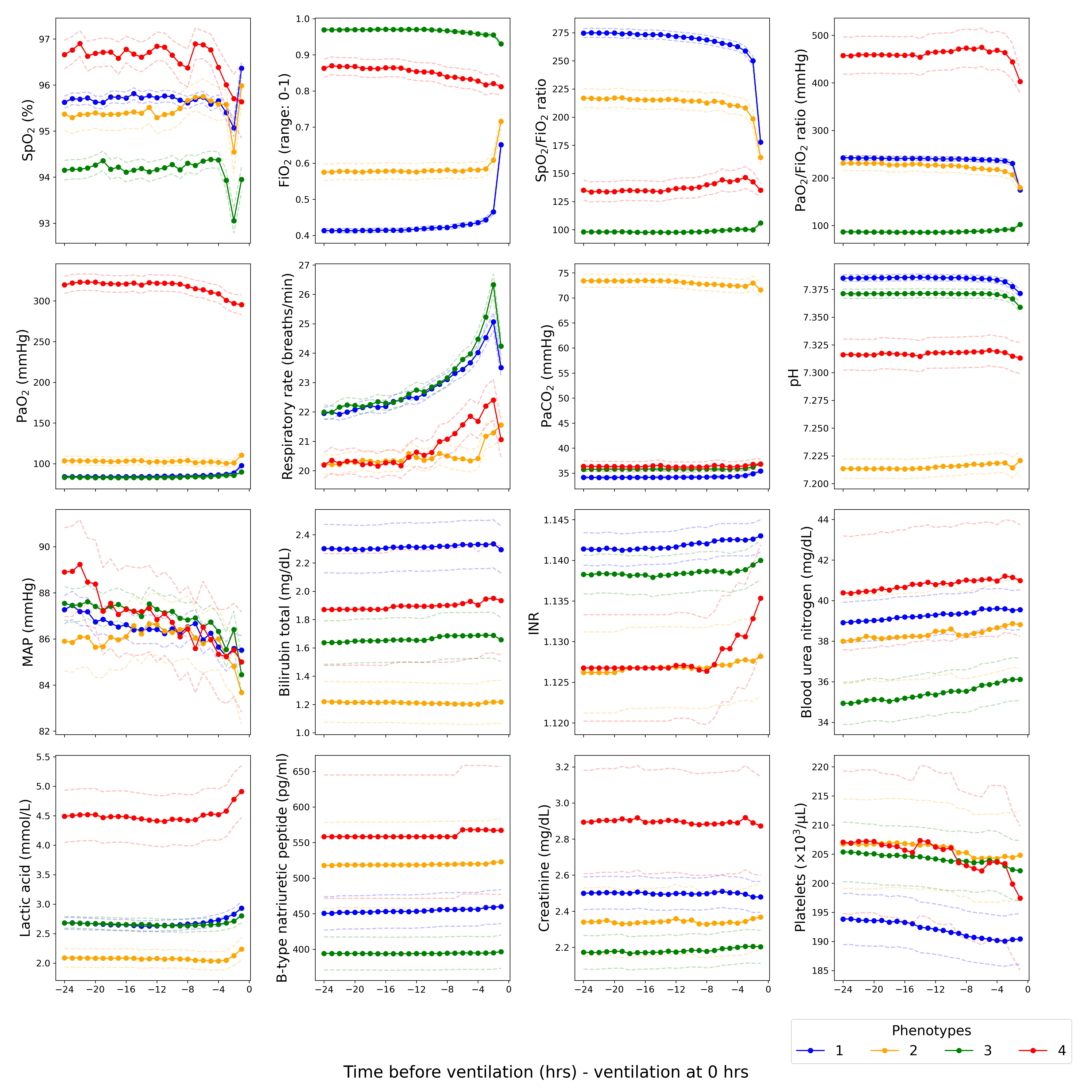}
    \caption{Trajectories of CRLI phenotypes over the 24-hour prior to ventilation with 83.4\% confidence intervals (dashed lines).}
    \label{fig:traj}
\end{figure*}

The mean trajectory for each phenotype was plotted for various indicators as shown in Fig.~\ref{fig:traj}. With dashed lines, $83.4\%$ confidence intervals were included in order to assess statistically significant differences ($p<0.05$) between points in the time series \cite{knol2011mis}. 

\begin{figure}[t]
    \centering
    \includegraphics[width=0.42\textwidth] {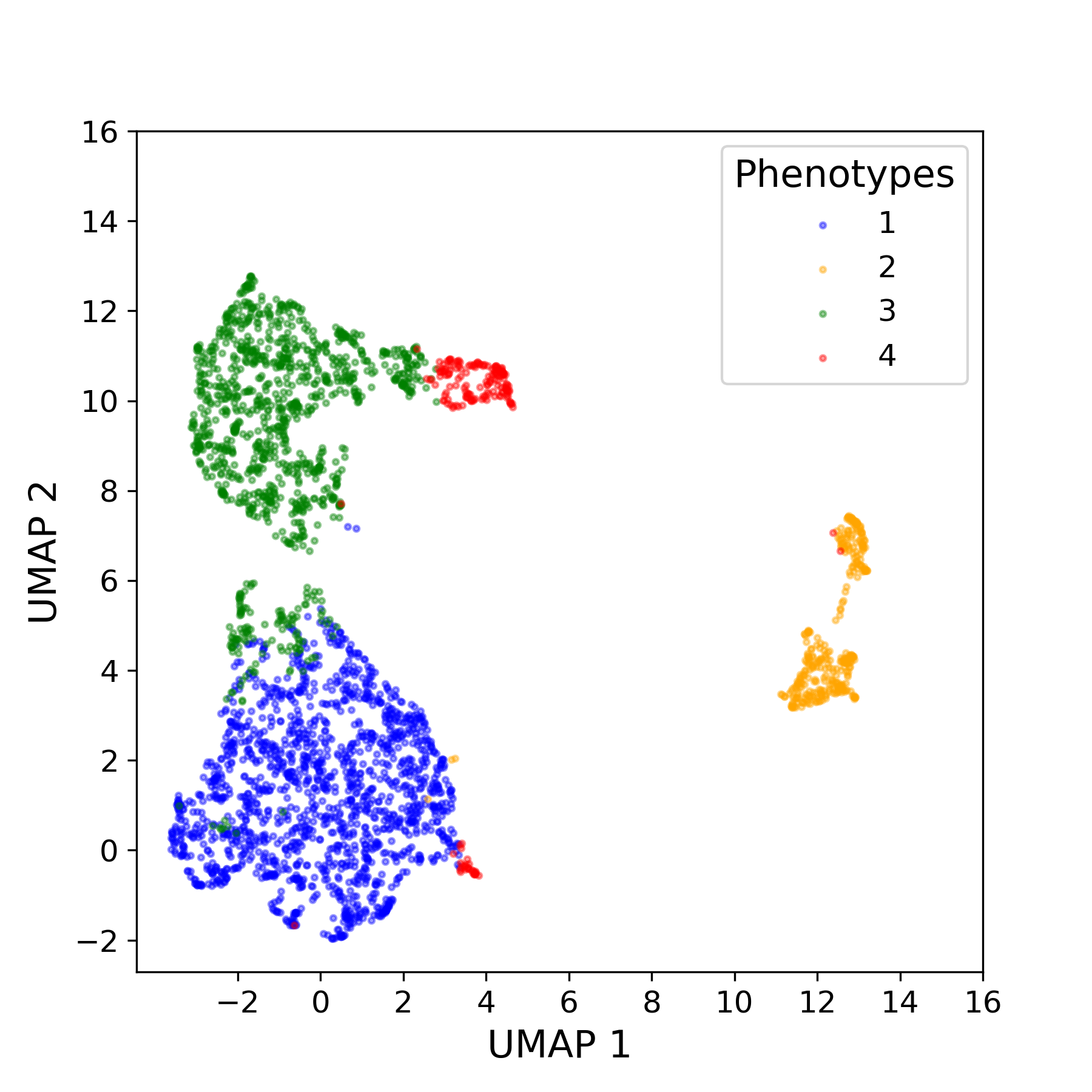}
    \caption{UMAP 2-D representation of clustering latent from CRLI, where identified phenotypes are highlighted in different color codes.}
    \label{fig:umap}
\end{figure}

For visualization of our phenotypes, Uniform Manifold Approximation and Projection (UMAP) \cite{mcinnes2018umap_published} scheme was applied on the latent representation (3349 $\times$ 128) to reduce the dimensions to (3349 $\times$ 2) as shown in Fig. ~\ref{fig:umap}. It is worth mentioning that UMAP was used solely for visualization of the phenotypes and not included in the clustering process itself.

\subsection{Kaplan-Meier Analysis}
\begin{figure}[t]
    \centering
    \includegraphics[width=0.45\textwidth] {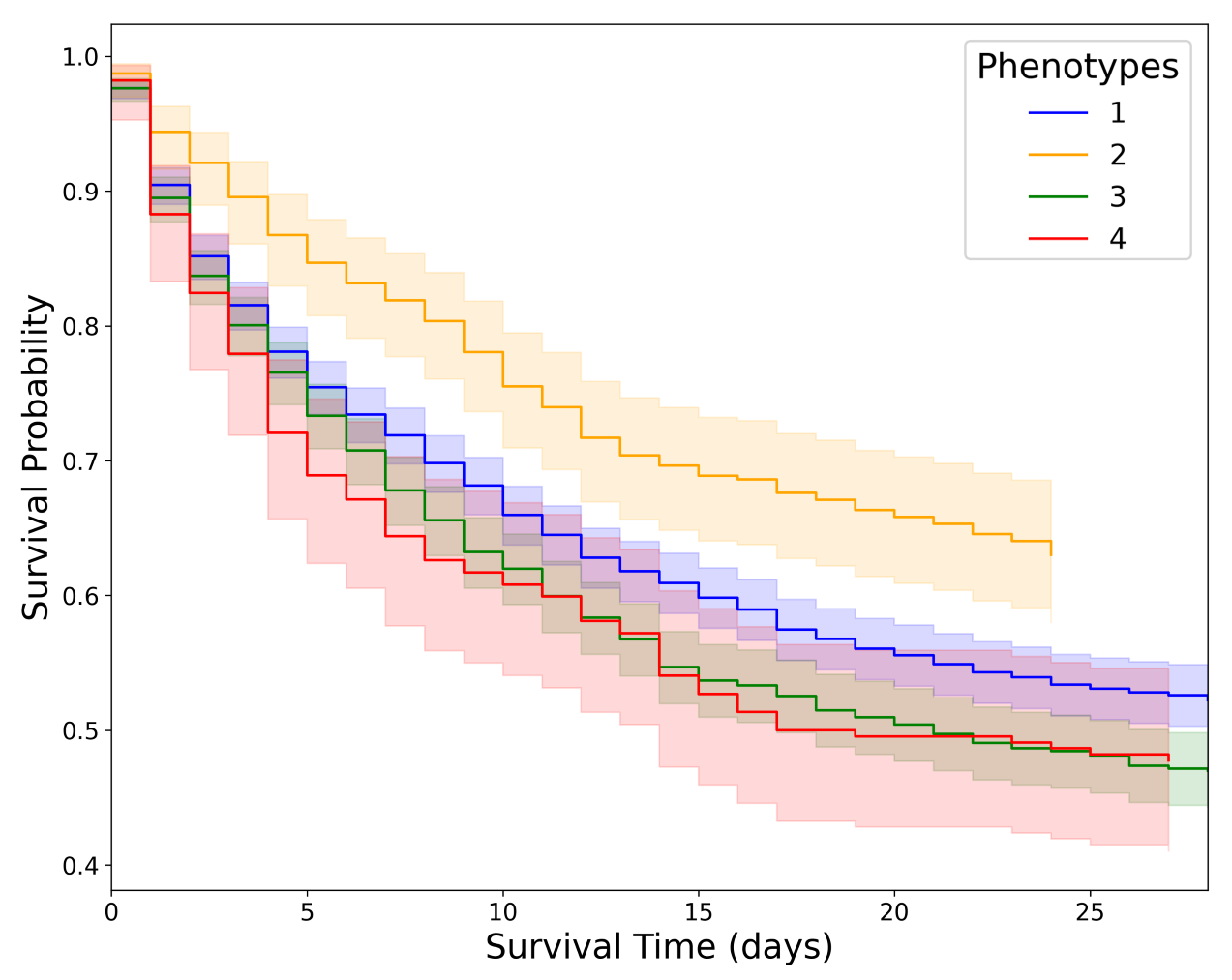}
    \caption{Kaplan-Meier curves on short-term mortality, showing survival probabilities for different phenotypes. A multivariate log-rank test on the survival curves was performed with $p<0.005$.}
    \label{fig:km}
\end{figure}

The Kaplan-Meier curves were plotted for the derived phenotypes to show the 28-day short-term survival probability as shown in Fig.~\ref{fig:km}. Phenotype 2 has the best survival probability trends followed by phenotype 1. Phenotype 3 and phenotype 4 have the worst survival probability trends.

\subsection{Comorbidity Analysis}
For each cluster, the Charlson Comorbidity Index (CCI) and the age-adjusted Charlson Comorbidity Index (ACCI) were calculated using ICD-10 codes \cite{charlson2022charlson}.  These comorbidity indexes estimate the risk of death with a higher number indicating a higher risk. All comorbidities were included at admission diagnosis. Charlson component scores were also calculated. These were a simple proportion of the patients within a phenotype that were affected by a component comorbidity from CCI. The python package \textit{comorbidipy} was used for the calculations\cite{pypiComorbidipy}. Both CCI and ACCI were computed with the Quan mappings and Quan weights. The CCI and ACCIs for various phenotypes are shown in Table~\ref{tab:cci}.

\begin{table}
{\centering
\caption{Demographics information of the four derived phenotypes.}

\scriptsize
\newcommand{\m}{\hphantom{$-$}}
\newcommand{\cc}[1]{\multicolumn{1}{c}{#1}}
\renewcommand{\tabcolsep}{0.37pc} % enlarge column spacing
\renewcommand{\arraystretch}{0.98} % enlarge line spacing

\begin{tabular}{ccccc}
\hline
         \textbf{Demo. variable}                                  &  \textbf{Phenotype 1}                             & \textbf{Phenotype 2}                             & \textbf{Phenotype 3}                             & \textbf{Phenotype 4}                            \\

\hline 
\#Encounters, n(\%)  & 1638 (48.91)  & 347 (10.36)  & 1169 (34.90)  & 195 (5.82)     \\
\#Patients, n   & 1586   & 332    & 1161     & 193    \\
%\#Fatalities, n   & 609   & 93   & 505  & 88 \\
Mortality, n(\%)  & 609 (38.40)   & 93 (28.01)     & 505 (43.50)     & 88 (45.60)      \\
Age, m(SD)             & 61.53 (15.27) & 64.51 (15.19) & 62.70 (15.89) & 62.53 (15.39) \\
Males, n(\%)                        & 882 (53.85)   & 176 (50.72)   & 641 (54.83)   & 115 (58.97)   \\
\hline
\textbf{Race} & & & & \\
Black, n(\%)   & 816 (49.82)   & 180 (51.87)   & 532 (45.51)   & 96 (49.23)    \\
White, n(\%)   & 676 (41.27)   & 159 (45.82)   & 519 (44.40)   & 88 (45.13)    \\
\hline 
\textbf{Ethnicity} & & & & \\
 Hispanic, n(\%)            & 62 (3.79)     & 10 (2.88)     & 53 (4.53)     & 4 (2.05)      \\
Non-Hisp., n(\%)         & 1443 (88.10)  & 321 (92.51)   & 1035 (88.54)  & 176 (90.26)  \\ 
\hline 
\end{tabular}}

\vspace{0.15cm}
{\scriptsize
Symbols used --- \#, n: count, m: mean, SD: standard deviation. Please note that mortality was computed with respect to patients (not encounters).} 
\label{tab:demographics}
\end{table}

\begin{table}[t]
\centering
\caption{Charlson comorbidity indexes (CCI) and age-adjusted CCI (ACCI) for the derived phenotypes on admission diagnosis.}
\begin{tabular}{|c|c|c|}
\hline 
\textbf{Phenotype} & \textbf{CCI} & \textbf{ACCI}  \\
\hline 
1                & 4.27         & 6.10 \\
2                & 4.11         & 6.17 \\
3                & 4.01         & 5.95 \\
4                & 4.30         & 6.17  \\
\hline
\end{tabular}
\label{tab:cci}
\end{table}
A one way ANOVA was performed on both CCI and ACCI. CCI was statistically significant with $p < 0.05$ while ACCI was statistically insignificant with $p = 0.42$. The frequency of certain pertinent comorbidities and Charlson components were compared between phenotypes for statistical significance using two proportion z-tests. Cardiac arrest rates were significantly higher for phenotype~4 at 37.44\% ($p < 0.00001$) when compared to all other phenotypes (phenotype~1 = 15.32\%, phenotype~2 = 20.46\%, phenotype~3 = 19.25\%). Charlson component score for chronic obstructive pulmonary disease was significantly higher for phenotype~2 at 0.5187 ($p < .00001$) when compared to all other phenotypes (phenotype~1 = 0.2851, phenotype~3 = 0.2797, phenotype~4 = 0.2769). Charlson component score for mild liver disease was significantly higher for phenotype~1 at 0.2735 when compared to all other phenotypes (phenotype~2 = 0.1470, $p < 0.00001$; phenotype 3 = 0.1993, $p < 0.00001$; phenotype 4 = 0.2000, $p < 0.05$). Charlson component score for moderate or severe liver disease was significantly higher for phenotype~1 at 0.1538 when compared to phenotype~2 and 3 but not phenotype~4 (phenotype~2 = 0.0634, $p < 0.00001$; phenotype~3 = 0.0924, $p < 0.00001$; phenotype~4 = 0.1282, $p = 0.34$).

\section{Discussion}
Four data-driven trajectory phenotypes were derived: a liver dysfunction/heterogeneous phenotype, a hypercapnia phenotype, a hypoxemia phenotype, and a MODS phenotype. Phenotype~1 is the a \textit{liver dysfunction/heterogeneous phenotype} with the fewest distinguishing features. It has a relatively average mortality of 38.4\%. It has the second highest CCI and the third highest ACCI. However, from the trajectory clustering, we note that phenotype 1 has a high International Normalized Ratio (INR), a high bilirubin total, and low platelets. This could indicate liver dysfunction and/or coagulopathy. Liver dysfunction is supported by the comorbidity analysis that suggests high incidence rates of mild liver disease and moderate to severe liver disease. It should also be noted that there is a precipitous drop in SpO$_2$/FiO$_2$ prior to ventilation. Patients in this phenotype seem to go from having little difficulty breathing to quickly being in a state of respiratory distress requiring mechanical ventilation similar to other phenotypes. 

Phenotype 2 is the \textit{hypercapnia phenotype} and has the lowest mortality of the phenotypes at 28.01\%. It also has the third highest CCI and tied for the highest ACCI. This phenotype has the lowest pH, the highest PaCO$_2$, and the highest levels of bicarbonate. These trajectories are suggestive of hypercapnia. From the comorbidity analysis, it is observed that phenotype 2 has the highest rates of chronic obstructive pulmonary disease (COPD) at 0.5187. COPD often leads to hypercapnia, making the high incidence rate unsurprising~\cite{csoma2022hypercapnia}. In patients with COPD, the ability to increase minute ventilation is limited, making their susceptibility to requiring mechnical ventilation a greater consequence of their baseline lung function rather than the severity of their septic process. This may explain the finding of having the lowest mortality among the phenotypes\cite{abdo2012oxygen}.

Phenotype 3 is the \textit{hypoxemia phenotype}. It has a relatively high mortality rate of 43.50\%, and also has the lowest CCI and ACCI. This phenotype has the highest FiO$_2$, lowest SpO$_2$/FiO$_2$ ratio, lowest PaO$_2$/FiO$_2$. These patients may benefit from subgroup analyzes, as they have the lowest burden of comorbid conditions, yet relatively high mortality and severe respiratory decline. Future studies that focus on this phenotype of patients may be able to identify targets for intervention prior to respiratory and other end-organ decline. 

Phenotype 4 is the \textit{multiple organ dysfunction syndrome (MODS) phenotype}. It has the highest mortality of the phenotypess at 45.60\%, and also has the highest CCI and ACCI. This phenotype has the highest creatinine, blood urea nitrogen (BUN), lactic acid, and b-type natriuretic peptide (BNP). However, the differences between phenotype 4 and the other phenotypes for BUN and BNP were not statistically significant. However, these differences suggest heart failure and kidney injury. Interestingly, phenotype 4 also had the highest SpO$_2$, PaO$_2$, and PaO$_2$/FiO$_2$ ratio indicating that these patients did not suffer from significant hypoxemia despite being the sickest phenotype.  From the comorbidity analysis, phenotype 4 had a significantly higher rate of cardiac arrest at 37.44\% as compared to the other phenotypes. The high mortality rate, high comorbidity indexes, trajectory phenotypes, and high cardiac arrest rate suggest MODS or multiple organ failure.

Some clinical significance can be derived from these phenotypes. While the distinction between hypoxemic and hypercapnic ARF translated into respective phenotypes, CRLI has yielded two additional phenotypes of liver dysfunction/heterogeneous phenotype and MOD phenotype. Of these, the liver dysfunction/heterogeneous phenotype and the hypercapnic respiratory failure phenotype appear to correlate with their underlying comorbidities. Interestingly, the hypoxemic ARF phenotype had the lowest CCI and ACCI, yet had the severest degree of hypoxemia and the second highest mortality. This phenotype identified by our clustering algorithm may represent a subgroup of sepsis-induced ARF patients who may experience the worst outcomes directly related to their respiratory failure, and may benefit from targeted interventions and adjunctive treatments for severe hypoxemia. The MOD phenotype, on the other hand, may represent a high-risk subgroup who may suffer ARF and poor outcomes as a consequence of multiple organ failures. Again, our clustering algorithm can help identify the sickest subset of patients with sepsis-induced ARF. The liver dysfunction/heterogeneous phenotype is interesting because it is the largest group, yet CLRI considered it as a separate phenotype and did not cluster it as one of the two main types of ARF. This suggests that classifying ARF simply as hypoxemic and hypercapnic respiratory failure may be an oversimplification and that additional comorbidities and organ failures may influence the clinical phenotype of sepsis-induced ARF.

The CRLI has been shown to be outperforming over earlier deep representation learning-based algorithms, such as VaDER, deep temporal clustering representation (DTCR), and deep temporal clustering (DTC) \cite{Ma_Chen_Li_Cottrell_2021}.
The technical capabilities of the state-of-the-art CRLI were shown in the results. By yielding phenotypes with statistically different Kaplan-Meier curves, producing a comparable number of clusters to the standard K-means + DTW algorithm, and reproducing two phenotypes that were already well established, CRLI demonstrated its ability to efficiently cluster our time series data. This also validates our decision of choosing the four phenotypes. In addition, multiple pre-processing and evaluation challenges were overcome. Multiple different imputation methods were utilized to handle the data with a high percentage of missing values. After creating the clusters, there was no obvious way to determine statistical significance between trajectories. Overlaying the 83.4\% confidence intervals was a simple-to-implement solution that allowed the viewer to see at what points the clusters were statistically significant from other clusters. This helped analysts and clinicians determine which trajectories had significant differences.

This study has several limitations. Our patient cohort is defined by the requirement for mechanical ventilation, which is a small but clinically important subset of patients with sepsis and septic shock. How these phenotypes apply \textit{a priori} to sepsis patients without respiratory failure, or sepsis patients who require non-invasive forms of respiratory support, is unclear. Furthermore, our work considered data only from a single academic hospital system, thus requiring further validation using comprehensive data from external centers to establish its generalizability and reproducibility across populations with different demographics and comorbidities. Additionally, sparse collection (not always in hourly intervals) of a few labs (e.g., PaO$_2$, PaCO$_2$) required imputation that sometimes rendered the trajectories flat. However, the value of these measurements usually do not change significantly over a short period of time. Finally, while using 83.4\% confidence intervals is a simple way to determine statistical significance, it does not take into consideration the shape of the trajectories. Trajectories that differ by shape may be missed if only the distance between them are assessed.

Despite these limitations, the developed ARF trajectory phenotyping study represents a pioneering effort to highlight distinct groups for MICU patients with a wide range of medical conditions.
We expect that a ML model trained on these derived phenotypes could be utilized prospectively in the ICU settings to get real-time phenotype predictions.
The next phases of this research will involve analyzing more patients' longitudinal data and conducting a prospective study for evaluating responses of different treatments, such as administration of high vs low PEEP strategy, vasopressors, and steroids.
We recommend that the results be validated by clustering data from other hospitals' MICUs or by trying to classify patients from other MICUs into one of the four phenotypes. Better data collection is also needed to reduce missing and spurious values. Finally, techniques to handle missing at random and missing not at random values should be explored \cite{ma2021identifiable}.

\section{Conclusion}
This exploratory study uses a multivariate set of features of six cardiorespiratory indicators and the new deep representation learning-based CRLI clustering algorithm to derive data-driven trajectory groups of patients with sepsis-induced ARF.
Our method yielded four unique and distinct groups that could be interpreted as four meaningful phenotypes. These phenotypes are clinically characterized as liver dysfunction/heterogeneous, hypercapnia, hypoxemia, and MODS. 
The primary categories of ARF, hypercapnia and hypoxemia were well detected by the CRLI model, and two other phenotypes were revealed that did not fit either category. 
With the comorbidity analysis and Kaplan-Meier analysis on mortality data, we provided substantial characteristics and differences in all derived clustered trajectories.
In addition to the clinically relevant results, we demonstrated the potential for the use of multivariate time series clustering algorithms in the exploration of ICU data. Future studies are required to reveal the effects of treatment on the derived phenotypes to optimize critical care and improve patient outcomes.

\section*{References}
\bibliography{biblio}

\end{document}